\documentclass[twoside,onecolumn,showkeys,aps]{revtex4}
\usepackage{eurosym}
\usepackage{bm}
\usepackage[utf8]{inputenc}
\usepackage{amssymb,theorem,xspace}
\usepackage{float}
\usepackage{amsmath}
\usepackage{mathrsfs}
\usepackage{graphicx}
\usepackage{subfigure}
\usepackage{soul}
\usepackage{xcolor}
\usepackage{natbib}

\begin{document}

\title{Fractional Effective Quark-Antiquark Interaction in Symplectic
Quantum Mechanics}
\author{M. Abu-Shady}
\email{dr.abushady@gmail.com}
\affiliation{Department of Mathematics and Computer Science, Faculty of Science, Menoufia
University, Shibin Al Kawm, Egypt}
\author{R. R. Luz}
\email{renato.luz@aluno.unb.br}
\affiliation{International Center of Physics, Instituto de Física, Universidade de
Brasília, 70.910-900, Brasília, DF, Brazil }
\author{G. X. A. Petronilo}
\email{gustavo.petronilo@aluno.unb.br}
\affiliation{International Center of Physics, Instituto de Física, Universidade de
Brasília, 70.910-900, Brasília, DF, Brazil }
\author{A. E. Santana}
\email{a.berti.santana@gmail.com}
\affiliation{International Center of Physics, Instituto de Física, Universidade de
Brasília, 70.910-900, Brasília, DF, Brazil }
\author{R. G. G. Amorim}
\email{ronniamorim@gmail.com}
\affiliation{International Center of Physics and Gama faculty, Universidade de
Brasília, 72.444-240, Brasília, DF, Brazil }
\affiliation{Canadian Quantum Research Center,\\
204-3002 32 Ave Vernon, BC V1T 2L7  Canada}

\begin{abstract}
Considering the formalism of Symplectic Quantum Mechanics, we investigate a two-dimensional non-relativistic strong interacting system, describing a bound heavy quark-antiquark state. The potential has a linear component that is analysed in the context of generalized fractional derivatives. For this purpose,  the Schrödinger equation in phase space is solved with the linear potential. The ground state solution is obtained and analyzed through the Wigner function  for the meson $c\overline{c}$.  One basic and fundamental result is that the fractional quantum  phase-space analysis gives rise to the confinement of quarks in the meson, consistent with experimental results.
\end{abstract}

\keywords {Phase space, Fractional calculus, quark interaction}
\maketitle

\section{Introduction}\label{sec:Introduction}
 Over the last decades strong interaction has been analyzed by different  approaches, including quantum chromodynamics (QCD) sum rules and lattice QCD, providing   quantitative and  qualitative characteristics of the hadronic matter~\cite{A_1}. In particular, systems such as quark-antiquark lead to interesting  descriptions and a quantitative test for QCD and for both the particle-physics standard model~\cite%
{AS_2,A_2,A_62,A_63,A_64,A_65,A_66,A_67,A_54,A_55,A_56,A_57,A_58,A_59}. In the case of a quarkonium, a popular approach considers the interaction between a quark-antiquark in a meson through a spatial (Euclidian) potential, such as $V(r)$, where $r$ is the distance between the quarks. The quantum nature of the state and the mass spectrum are  studied by  considering as a model for the state time-evolution the Schr\"odinger equation. This corresponds to a specific sector of the strong interaction, which is called non-relativistic QCD.

From  the spectral analysis, it appears that the interaction in such  systems as a heavy quark and an antiquark (charmonium $c\overline{c}$) is successfully modeled by the Cornell potential, which defined by
\begin{equation}
 V(r) =\lambda r-\frac{\sigma}{r},\label{cornell_pt}
\end{equation}
where $\lambda$ is the string tension and $\sigma$ is the strong coupling constant. The linear term is associated with  the confinement, while the Coulomb-like term is a consequence of the asymptotic freedom. This potential has been used to investigate, as an instance, the confinement/deconfinement phase transition  in hadronic  matter~\cite{A_0,A_5,A_6,A_7,B_5}. In addition,  the Schr\"odinger equation with the Cornell potential as a model has been extensively used to explore quarkonium systems in the configuration space. This is the case of investigations of the heavy quarkonia mass spectroscopy and the bound state properties of $c\overline{c}$ and $b\overline{b}$
mesons~\cite{A_3,A_4,A_5,A_8,A_10,A_23,A_60,A_71,A_72}.

Considering theoretical aspects, the  heavy quarkonium characteristics have
been analyzed by the Schrödinger equation with the Cornell potential
through variational method in the framework of supersymmetric quantum
mechanics~\cite{A_6, A_9, A_11, A_12}. The
mass spectra of heavy quarks $b\overline{b}$, $c\overline{c}$ and $b%
\overline{c}$ within the framework of the Schrödinger equation with a
general polynomial potential were also addressed~\cite{A_1}. In this formulation, the Nikiforov-Uvarov (NU) method~\cite{A_6} was  used to
calculate  energy eigenvalues. The radial Schrödinger equation is extended to higher
dimensions, and the NU method is applied to a  Cornell-type
potential. As a consequence, in order to obtain the heavy quarkonium masses, the energy
eigenvalues and the associated wave functions are determined~\cite{AS_2}. The eigen-solutions and an inverted polynomial potential were obtained by using the
NU procedure~\cite{A_11, A_12,A_13, RL_1}.

It is important to notice that although the Cornell potential presents theoretical and experimental consistency with the standard-model, aspects of the hadronic matter as confinement are not obviously derived from the model-dependent Schr\"odinger equation. Indeed this is the case of  solutions of the Schr\"odinger equation in the Euclidian space representation. However, the recent analysis of the Wigner function of such a system as a $c\overline{c}$ mesons provides an interesting description of the confinement, by using the characteristics of the phase-space quantum mechanics.  These achievements were carried out by considering the symplectic quantum mechanics, in which the Schr\"odinger equation is written in a phase-space representation~\cite{RL_1}. Nevertheless, accounting for the physical richness of phase space, many aspects remain to be explored, such as the fractional structure of the symplectic Schr\"odinger describing a quark-antiquark system.

The use of fractional calculus has attracted   attention  in  a variety of areas in physics~\cite%
{AS_1,AS_3,B_1,B_2,B_3,mh_1}. For heavy quarkonium systems, methods as the NU formulation and analytical iteration have been explores to provide analytical solutions of the  N-dimensional radial Schr\"odinger  equation in the framework of fractional
space~\cite{AS_3,AS_4}. A category of potentials including the oscillator potential, Woods-Saxon potential, and the Hulthen potential, have also been studied analytically with fractional radial Schrödinger equation by NU method~\cite{AS_4,B_4}. In order to investigate the binding energy and temperature dissociation, the conformable fractional formulation was extended to a finite temperature context~\citep{AS_4}. A fundamental goal of the present work is to survey the applicability of the fractional approach to the study of quark dynamics in phase space.

Then the behavior of the Wigner function for the ground state of $c\overline{c}$ meson is analyzed from the perspective of fractional calculus. For this purpose, the symplectic Schr\"odinger equation is rewritten in the fractional form with the linear term of Cornell potential for the heavy $c\overline{c}$ meson. Beyond physical aspects, the analysis provides a simpler procedure to study this type of systems.

The work is organized as follows. In Section~\ref{sqm}, some aspects of the Schr\"o%
dinger equation represented in phase space are reviewed in particular to fix the notation. In Section~\ref%
{linear}, the concept of fractional derivative is implemented in the
symplectic Schr\"odinger equation for the linear part of the Cornell potential. Section~\ref{DR} is devoted the discussion of outcomes. In Section~\ref{FR},
summary and final concluding remarks are presented.

\section{Symplectic Quantum Mechanics: notation and Wigner function}
\label{sqm}

 Considering a phase space manifold $\Gamma$, where a point is specified by a set of Real coordinates $(q,p)$, the complex valued square-integrable
functions, $\phi (q,p) \in \Gamma $, such that, $\int dpdq\;\phi ^{\ast
}(q,p)\phi (q,p)<\infty $,  is equipped with a Hilbert
space structure, $\mathcal{H}(\Gamma)$. Here  $q$ stands by a vector in the $\mathbb{R}^3$ Euclidian manifold, and $p$ stands for points in the dual $\mathbb{R^*}^3$. The point $(q,p)$ is a vector in the cotangent-bundle of  $\mathbb{R}^3$, equipped with a symplectic two-form ~\cite{A_4}.  In this way, $(q,p)$ can be used to introduce a basis in $\mathcal{H}(\Gamma)$, denoted by $|q,p\rangle $ with
completeness relation given by $\int dpdq|q,p\rangle \langle q,p|=1$. It follows that $%
\phi (q,p)=\langle q,p|\phi \rangle $, where $\langle \phi |$ is the dual
vector of $|\phi \rangle $. The symplectic
Hilbert space  $H(\Gamma)$ can be used as the representation space of symmetries. For the non-relativistic Galilei group,  position and momentum
operators are written as
\begin{eqnarray}
\widehat{P} &=&p\star =p-i\frac{\partial }{\partial q},  \label{g1} \\
\widehat{Q} &=&p\star =q+i\frac{\partial }{\partial p}.  \label{g2}
\end{eqnarray}%
A symplectic structure of quantum mechanics is constructed in the following way. The
Heisenberg commutation relation $\left[ \widehat{Q},\widehat{P}\right] =i$
is fulfilled. And then using the following operators
\begin{eqnarray}
\widehat{K}_{i} &=&m\widehat{Q}_{i}-t\widehat{P}_{i}, \\
\widehat{L}_{i} &=&\epsilon _{ijk}\widehat{Q}_{j}\widehat{P}_{k}, \\
\widehat{H} &=&\frac{\widehat{P}^{2}}{2m},
\end{eqnarray}%
the set of commutation rules are obtained
\begin{equation}
\begin{aligned}
&\left[\widehat{L}_i,\widehat{L}_j\right]=i\epsilon_{ijk}\widehat{L}_k,%
\qquad\left[\widehat{L}_i,\widehat{k}_j\right]=i\epsilon_{ijk}\widehat{K}_k,
\qquad \left[\widehat{K}_i,\widehat{H}\right]=i\widehat{P}_i,\\
&\left[\widehat{L}_i,\widehat{P}_j\right]=i\epsilon_{ijk}\widehat{P}_k,%
\qquad\left[\widehat{K}_i,\widehat{P}_j\right]=i m\delta_{ij}\mathbf{1},\\
\end{aligned}
\end{equation}
being zero for all the other commutations. It is known as Galilei-Lie
algebra and $m$ is a central extension. The Galilei symmetries are  defined by the
operators $\widehat{P},\widehat{K},\widehat{L}$ and $\widehat{H}$, which stands, respectively, by the
generators of spatial translations, Galilean boosts, rotations and time translations.

The time-translation generator, $\widehat{H}$, leads to the time evolution of a symplectic wave function, i.e.,
\begin{equation}
\psi(q,p,t)=e^{\widehat{H}t}\psi(q,p,0).
\end{equation}
The infinitesimal version of this equation reads as
\begin{eqnarray}
\partial_t \psi(q,p;t)&=&\widehat{H}(q,p) \psi(q,p;t), \\
\end{eqnarray}
the Schrodinger-type equation in $\Gamma$~\cite{A_33}.

The physical interpretation of this formalism is obtained by the association of $\psi(q,p,t)$ with a function $f_W$, i.e.,~\cite{A_70,A_37,A_38}
\begin{equation}
f_W(q,p,t)=\psi(q,p,t)\star\psi^\dagger(q,p,t),
\end{equation}
 In
the next section, the representation symplectic Schrödinger equation
in fractional context for the heavy quark system $c\overline{c}$   is obtained.

\section{Fractional symplectic Schrödinger equation for the confinement potential}

\label{linear} In this section, the symplectic Schrödinger equation is
generalized to a fractional-space Schrödinger equation describing two particles
interacting to each other by the linear part of Cornell potential. Using the results of the previous section, the symplectic Schrödinger equation
takes the form~\cite{RL_1}
\begin{equation}
\frac{(p\star)^{2}}{2m} \psi (q,p)+\lambda (q\star) \psi (q,p)=E\psi (q,p).
\label{D_0}
\end{equation}
Using the Eqs.~(\ref{g1}) and (\ref{g2}) in (\ref{D_0}), it leads to
\begin{equation}
\frac{1}{2m}\left( p^{2}-ip\partial _{q}-\frac{1}{4}\partial _{q}^{2}\right)
\psi +\lambda \left( q+\frac{i}{2}\partial _{p}\right) \psi =E\psi ,
\label{D_3}
\end{equation}%
where natural units are used, such that $\hbar =1$. By using the transformation
$\omega =\frac{p^{2}}{2m}+\lambda q$, this equations reads
\begin{equation}
\frac{\partial ^{2}\psi (\omega )}{\partial \omega ^{2}}-\frac{\omega -E}{%
\eta }\psi (\omega )=0.  \label{S_2}
\end{equation}%
where $\eta =\frac{\lambda ^{2}}{8m}$. Writing Eq.~\eqref{S_2} in fractional
from~\cite{AS_1}, it follows that
\begin{equation}
\frac{1}{{\varsigma }^{2(1-\alpha )}}D^{\alpha }D^{\alpha }\psi (\omega )-\frac{%
\omega -E}{\eta }\psi (\omega )=0,
\end{equation}%
where
\begin{equation}
D^{\alpha }\psi (\omega )=\frac{\Gamma (\beta )}{\Gamma (\beta -\alpha +1)}%
\omega ^{1-\alpha }\frac{\partial \psi }{\partial \omega },
\end{equation}%
and $\varsigma $ is a scalar factor, $0<\alpha \leq 1$ and $0<\beta \leq 1$.
Thus
\begin{equation}
D^{\alpha }D^{\alpha }\psi (\omega )=\bigg(\frac{\Gamma (\beta )}{\Gamma
(\beta -\alpha +1)}\bigg)^{2}\bigg[\omega ^{2-2\alpha }\frac{d^{2}\psi }{%
d\omega ^{2}}+(1-\alpha )\omega ^{1-2\alpha }\frac{d\psi }{d\omega }\bigg],
\end{equation}%
Therefore, Eq.~\eqref{S_2} in the fractional form is written as
\begin{equation}
\frac{d^{2}\psi }{d\omega ^{2}}+\frac{(1-\alpha )}{\omega }\frac{d\psi }{%
d\omega }-\bigg(\frac{\omega -E}{A\eta }\bigg)\omega ^{2\alpha -2}\psi =0,
\label{frac}
\end{equation}%
where
\begin{equation}
A=\frac{1}{\varsigma ^{2(1-\alpha )}}\bigg(\frac{\Gamma (\beta )}{\Gamma (\beta
-\alpha +1)}\bigg)^{2},  \label{S_3}
\end{equation}%
with $\varsigma $ being a scale factor. It worth noting that if $\alpha =\beta =1$ one obtains
the original Eq.~\eqref{S_2}.

\section{Discussion of Results}

\label{DR}

In this section, in order to obtain an analytical function of Eq.~\eqref{frac}, the fractional
parameter is taken as $\alpha=0.5$ (For other values,  perturbative methods can be used. This will be not addressed in the present paper). This leads to
following form
\begin{equation}
\frac{d^2\psi}{d\omega^2}+\frac{1}{2\omega}\frac{d \psi}{d \omega}-\bigg(%
\frac{\omega-E}{A\eta\omega}\bigg)\psi=0.  \label{frac_0.5}
\end{equation}
The solution of this equation is given by
\begin{equation}
\psi= C_1\sqrt{\omega}\,e^{-\frac{\omega}{\kappa}}\text{M}\Bigg(-\frac{-3\sqrt{A\eta} + 2E}{%
4\sqrt{A\eta}}, 3/2, \frac{2\omega}{\sqrt{A\eta}}\Bigg) + C_2\sqrt{\omega}%
\,e^{-\frac{\omega}{\kappa}}\text{U}\Bigg(-\frac{-3\sqrt{A\eta} + 2E}{4\sqrt{A\eta}}, 3/2,
\frac{2\omega}{\sqrt{A\eta}}\Bigg).
\end{equation}
where $C_1$, $C_2$ are constants and $\text{M}(a, b, z)$, $\text{U}(a, b, z)$
are the Kummer functions. One can regard U$(a, b, z)$ as a physical solution
since it is the only one that is square integrable. As a result, one can
impose that $C_1 = 0$. Additionally, if $a = -n$, the series U$(a, b,z)$
becomes a polynomial in $\omega$ of degree not exceeding $n$, where $n = 0, 1,
2,...$ This circumstance allows us to write
\begin{equation}
\psi_n(\omega)=C_n\sqrt{\omega}\,e^{-\frac{\omega}{\kappa}}\text{U}\Bigg(-n, 3/2, \frac{%
2\omega}{\kappa}\Bigg).
\end{equation}
where $\kappa=\sqrt{A\eta}$, and
\begin{equation}\label{energy1}
E_n=\kappa\bigg(2n+\frac{3}{2}\bigg).
\end{equation}
Notice that the energy does not depend explicitly on the kinetic energy, thus the initial
condition should be $q=p=0$.

For the ground state, making the substitution into $q$ and $p$ again, one
have
\begin{equation}
\psi_0(q,p)=C_0\sqrt{\frac{p^2}{2m}+\lambda q\,}\;\exp\bigg(-\frac{{\frac{p^2%
}{2m}+\lambda q}}{\kappa}\bigg),
\end{equation}
and
\begin{equation}
E_0=\kappa\bigg(\frac{3}{2}\bigg).
\end{equation}

Using the fact that $\psi(q,p)$ is real, the normalized Wigner function of
the ground state is given by
\begin{equation}
f_{W_0}=\psi_0\star\psi_0=C_0^2(\kappa)\;\sqrt{\frac{p^2}{2m}+\lambda q\,}\;\exp\bigg(-\frac{{\frac{p^2%
}{2m}+\lambda q}}{\kappa}\bigg),\label{eq_19}
\end{equation}
where the constant $C_0(\kappa)$ depends on the value of $\kappa$.

\begin{figure}[th!]
\centering
\includegraphics[scale=0.28]{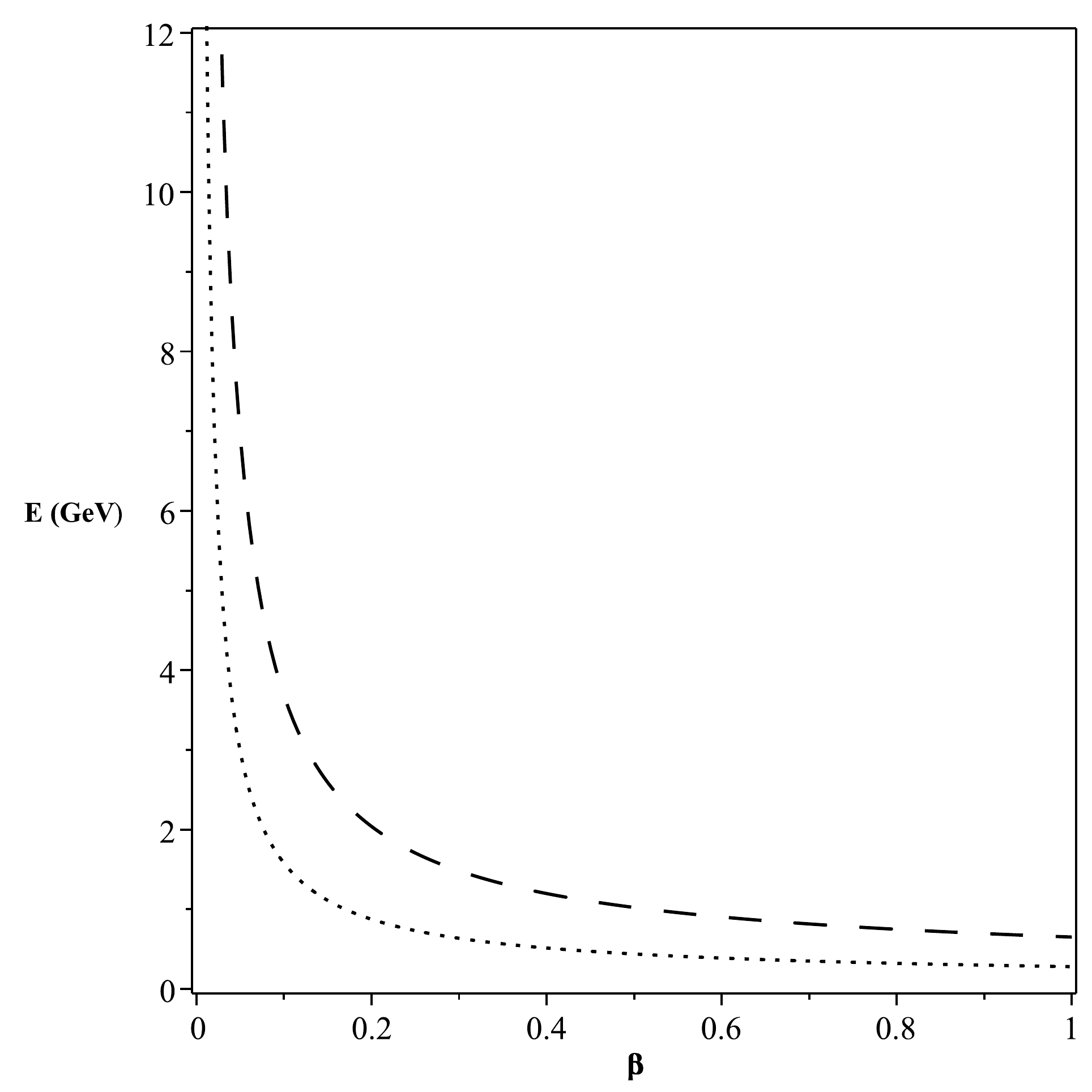}
\caption{ The dot-line (......) stands for the fundamental level of energy, $E_0$, and the dashed-line (- - - - -) represents the first exited energy level, $E_1$; both as a function of the parameter $\beta$, with  $\varsigma = 1.$ The mass of the quarks are taken as $m = 0.336 $ $MeV$ and $\lambda = 0.22$  $MeV^2$~\cite{corn333,cor334}.}
\label{esp1}
\end{figure}

\begin{figure}[th!]
\centering
\includegraphics[scale=0.28]{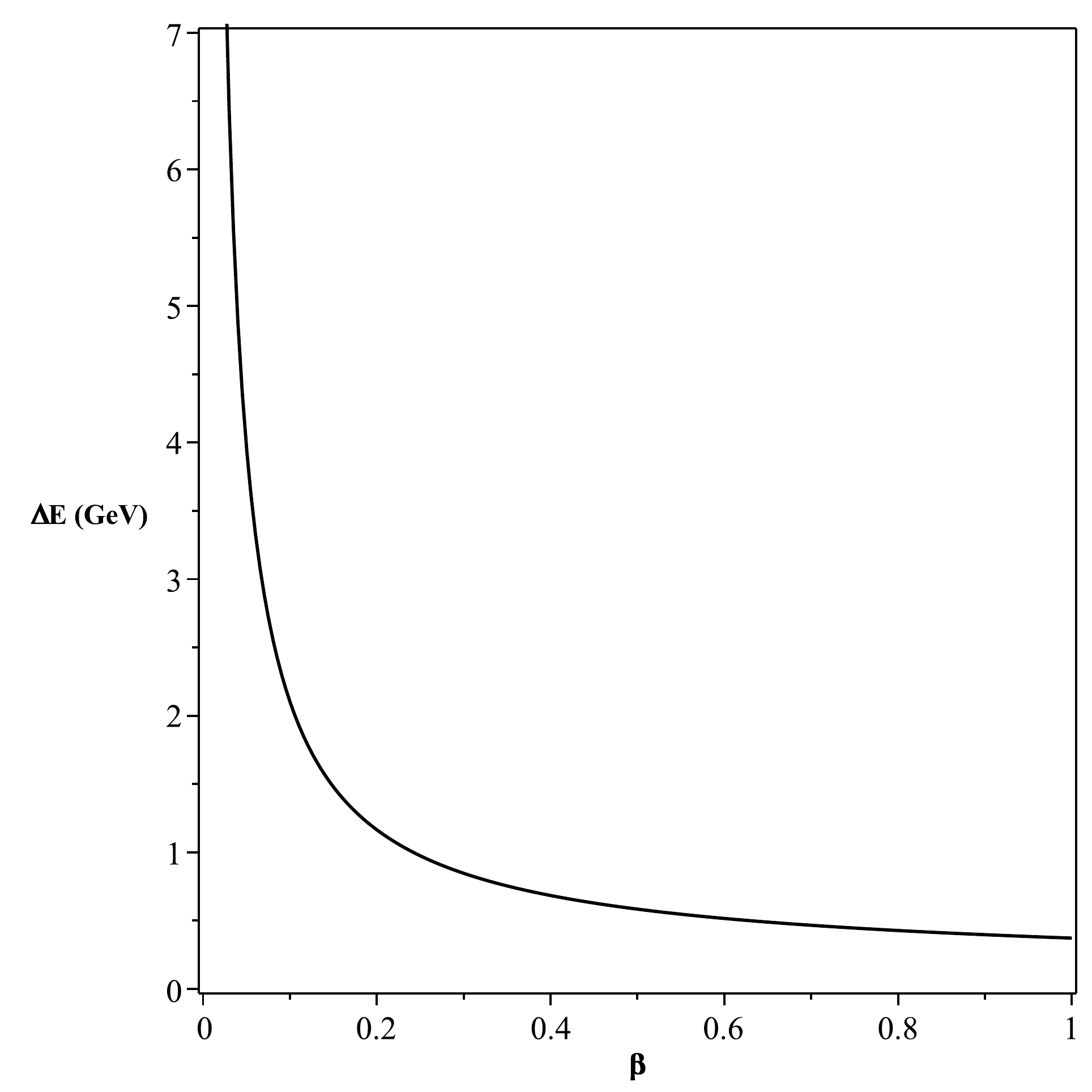}
\caption{ The difference of the levels of energy $\Delta E=E_1 - E_0$ as a function of  the parameter $\beta$, with  $\varsigma = 1.$
 } 
  \label{esp2}
\end{figure}

In Fig.~\ref{esp1}, the behaviour of the $E_0= E_0(\beta)$ and $E_1= E_1(\beta)$ is described. In Fig.~\ref{esp2}, the difference $\Delta E=E_1 - E_0$ is plotted as a function of the parameter $\beta$, considering $\zeta = 1$. This difference has for $\beta \approx 0.3$ reached the order of value of experimental measurements~\cite{corn333}. It is worth emphasizing here that the linear part of the Cornell potential only does not provide a spectrum in agreement with experimental measurement~\cite{corn333,cor334}. Here since we have the parameters of the fractional derivatives, those results can be improved for  values of $\zeta$ and $\beta$. The next point is to explore the behavior of the Wigner function in order to detail the behaviour of the confinement of quark-antiquark.

\begin{figure}[H]
\centering
\includegraphics[clip, trim=0.9cm 8.5cm 0.9cm 8.5cm, width=12cm]
{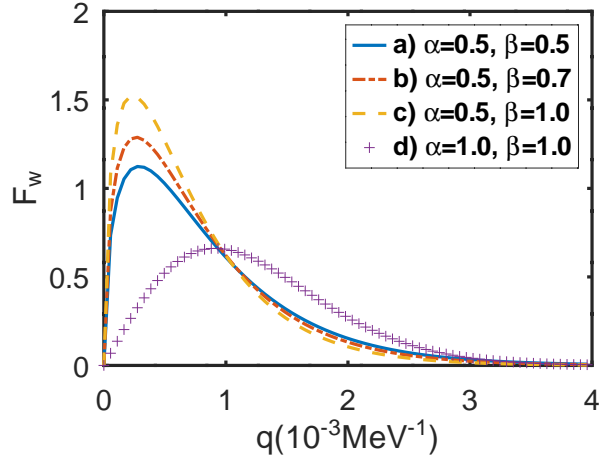}
\caption{Comparison of fractionary Wigner functions and the non-fractionary
one in the cut off of $\dfrac{p^2}{2m}=0$ in Eq.~\eqref{eq_19}. Increasing the value of the $\dfrac{p^2}{2m}$ in the cut of the Wigner function will show a shift for the left, thus giving a maximum value of $p$. for the existence of the function. The original Wigner function is given when $\alpha=\beta=1$. }
\label{fig2}
\end{figure}

  Wigner function for the  fractional parameter for $\alpha=0.5$, and different values of $\protect\beta $ is presented in Fig.~\ref{fig2}. The figure compares fractional Wigner functions to the original one, which is $\alpha =\beta =1$. We observe that the peaks diminish by lowering $\beta$.

\begin{figure}[H]
\centering
\includegraphics[clip, trim=0.9cm 8.5cm 0.9cm 8.5cm, width=12cm]
{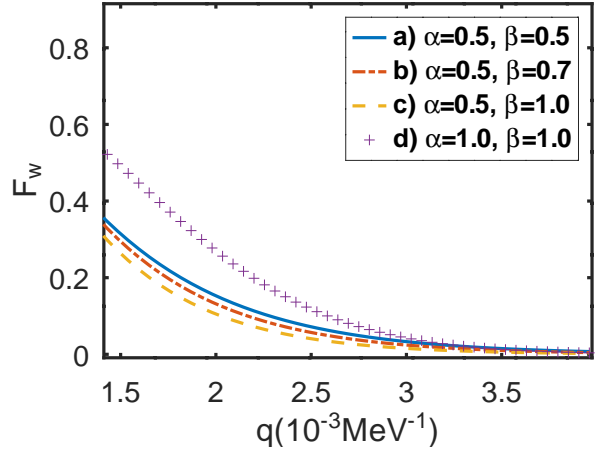}
\caption{It is presented only a small part of Fig.~\ref{fig2}. It is observed that the maximum distance of existence decreases with lowering $\beta$.}
\label{fig3}
\end{figure}

The curves a) to c) of Fig.~\ref{fig2} shows that increasing $\beta$ the peaks of Wigner functions increase. Additionally, we see that the peaks decrease to zero as $\beta$ goes to zero. Does $\beta$ functions as a fitting parameter for the fractional Wigner function of the $c\bar{c}$ meson. The curve d) is the original Wigner function without the fractional parameters for $c\bar{c}$ meson~\cite{AS_1}. In Fig.~\ref{fig3} shows that increasing $\beta$ the distance $q$ decreases. The maximum value of $\beta$ is the best fit for the case of $\alpha=0.5$. When compared to the experimental evidence; for comparison, the experimental value for the maximum distance is $q_0=4.077\cdot10^{-3}MeV^{-1}$~\cite{A_1}.

For the case of general Cornell potential, Eq.~\eqref{cornell_pt},
one can linearize to get an approximation form. In the first approximation
\begin{equation}
V(q)=-\frac{2\sigma}{q_0}+\Big(\lambda+\frac{\sigma}{q_0^2}\Big)q,
\end{equation}
where $a,b$ and $q_0$ are constants.
Then the Hamiltonian is
\begin{equation}
    (H\star)\psi=\frac{(p\star)^2}{2m}\psi +\Big(\lambda+\frac{\sigma}{q_0^2}\Big)(q\star)\psi
    =\Big(E+\frac{2\sigma}{q_0}\Big)\psi.\label{eq_cp}
\end{equation}
This equation leads to
\begin{equation}
    (H\star)\psi=\frac{(p\star)^2}{2m}\psi +\lambda^\prime (q\star)\psi=E^\prime\psi,\label{eq_cp1}
\end{equation}
It is worth noting that Eq.~\eqref{eq_cp1} is the same as Eq.~\eqref{D_0} with $\lambda^\prime=\lambda+\dfrac{\sigma}{q_0^2}$ and $E^\prime=E+\frac{2\sigma}{q_0}$. Therefore the same analysis applies here. The energy is given by
\begin{equation}
    E_n=\kappa\bigg(2n+\frac{3}{2}\bigg)-\frac{2\sigma}{q_0},
\end{equation}
where $\kappa=\sqrt{A\eta}$ and $\eta=\frac{{\lambda^\prime}^2}{8m}$. It is noteworthy that, when $q\rightarrow 0$, in the general Cornell potential has the $q^{-1}$ that is responsible by interaction at short distances and corresponding to one gluon exchange. In addition, in Table 1 presents the theoretical results from the fractional model for $\alpha=0.5$, $\beta=1.0$, calculated from Eq.(\ref{energy1}), and the respective experimental values. Comparisons were established only for 1S states, as our theoretical model is applicable only to such states. We didn't include spin in our theoretical model. We noticed that there is good accuracy between the theoretical and experimental results \cite{particle}, better than those obtained by other theoretical models \cite{corn333}. 
\begin{table}[h!]
    \centering
\caption{Experimental and Theoretical masses (in $MeV$) of charmonium mesons.}
\begin{tabular}{c c c }
\hline
Meson & Fractional Cornell Potential & Experimental Data \cite{particle} \\
\hline
$J/ \Psi(1S)$ & $3,1003$ & $3,0969$\\
$\Upsilon(1S)$ & $9,4818$ & $9,4603$\\
$\eta(1S)$ & $2,7992$ & $2,9796$ \\
\hline
\end{tabular}
\end{table}

\section{Final Remarks}
We have studied the symplectic Schr\"odinger-like equation in the presence of a linear potential using the formalism of generalized fractional derivatives for non-relativistic heavy quarkonium bound state. For this purpose we have investigated the behavior of the Wigner function for the ground-state $c\overline{c}$ meson considering  the symplectic quantum mechanics and the generalized fractional derivative constructed in~\cite{RL_1,AS_1}. The Wigner function has been obtained for the charmonium state in the fractional form using the generalized fractional derivative as in Ref.~\cite{B_1}, where we obtained the classical case at $\alpha =\beta =1$. To obtain an analytical solution, we analyse the case  $\alpha=0.5$ . For this value of $\alpha$, it was observed that the peaks of the Wigner function is lowered by decreasing fractional parameter $\beta$, therefore, this parameter can be used as a fitting parameter. For the case of $\alpha=0.5$ the value $\beta=1$ is the best fit considering the experimental evidence. Therefore, the present analysis, seems to indicate the relevance of such a generalized fractional model based on the symplectic Schrödinger equation with linear term (Cornell potential) as far as quarkonium dynamics in phase space is concerned. To further the study of a quarkonium system within the fractional and phase space approaches, we will  include a quadratic term (or correction term) at the Cornell potential and other values for the fractional parameter $\alpha$. We also intend to study spinorial systems. 
\vskip 16pt
\label{FR}
\vskip 16pt
\textbf{Acknowledgments}

This work was supported by CNPq and CAPES (Brazilian Agencies).

\end{document}